\documentclass[twocolumn]{aastex631}

\usepackage{color, booktabs, subfigure, multirow}
\usepackage{url, hyperref,amsmath}
\usepackage[capitalise]{cleveref}
\usepackage[hang,flushmargin]{footmisc}

\def\msh15{{MSH~15$-$5{\sl 2}}}

\shorttitle{X-ray polarization of \msh15}
\shortauthors{Romani et al.}

\graphicspath{{./}}

\begin{document}

\title{The Polarized Cosmic Hand: {\it IXPE} Observations of PSR B1509-58/\msh15}

\correspondingauthor{Roger W. Romani}
\email{rwr@astro.stanford.edu}
\author[0000-0001-6711-3286]{Roger W. Romani}
\affiliation{Department of Physics and Kavli Institute for Particle Astrophysics and Cosmology, Stanford University, Stanford, California 94305, USA}

\author[0000-0001-6395-2066]{Josephine Wong}
\affiliation{Department of Physics and Kavli Institute for Particle Astrophysics and Cosmology, Stanford University, Stanford, California 94305, USA}

\author[0000-0002-7574-1298]{Niccol\'{o} Di Lalla}
\affiliation{Department of Physics and Kavli Institute for Particle Astrophysics and Cosmology, Stanford University, Stanford, California 94305, USA}

\author[0000-0002-5448-7577]{Nicola Omodei}
\affiliation{Department of Physics and Kavli Institute for Particle Astrophysics and Cosmology, Stanford University, Stanford, California 94305, USA}

\author[0000-0002-0105-5826]{Fei Xie}
\affiliation{Guangxi Key Laboratory for Relativistic Astrophysics, School of Physical Science and Technology, Guangxi University, Nanning 530004, China}
\affiliation{INAF Istituto di Astrofisica e Planetologia Spaziali, Via del Fosso del Cavaliere 100, 00133 Roma, Italy}

\author[0000-0002-5847-2612]{C.-Y. Ng}
\affiliation{Department of Physics, The University of Hong Kong, Pokfulam, Hong Kong}

\author[0000-0003-1074-8605]{Riccardo Ferrazzoli} 
\affiliation{INAF Istituto di Astrofisica e Planetologia Spaziali, Via del Fosso del Cavaliere 100, 00133 Roma, Italy}

\author[0000-0003-0331-3259]{Alessandro Di Marco}
\affiliation{INAF Istituto di Astrofisica e Planetologia Spaziali, Via del Fosso del Cavaliere 100, 00133 Roma, Italy}

\author[0000-0002-8848-1392]{Niccol\'{o} Bucciantini}
\affiliation{INAF Osservatorio Astrofisico di Arcetri, Largo Enrico Fermi 5, 50125 Firenze, Italy}
\affiliation{Dipartimento di Fisica e Astronomia, Universit\`{a} degli Studi di Firenze, Via Sansone 1, 50019 Sesto Fiorentino (FI), Italy}
\affiliation{Istituto Nazionale di Fisica Nucleare, Sezione di Firenze, Via Sansone 1, 50019 Sesto Fiorentino (FI), Italy}

\author[0000-0001-7397-8091]{Maura Pilia}
\affiliation{INAF Osservatorio Astronomico di Cagliari, Via della Scienza 5, 09047 Selargius (CA), Italy}

% ...

\begin{abstract}
We describe {\it IXPE} polarization observations of the Pulsar Wind Nebula (PWN) \msh15, the `Cosmic Hand'. We find X-ray polarization across the PWN, with B field vectors generally aligned with filamentary X-ray structures. High significance polarization is seen in arcs surrounding the pulsar and toward the end of the `jet', with  polarization degree $PD>70\%$, thus approaching the maximum allowed synchrotron value. In contrast, the base of the jet has lower polarization, indicating a complex magnetic field at significant angle to the jet axis. We also detect significant polarization from PSR B1509$-$58 itself. Although only the central pulse-phase bin of the pulse has high individual significance, flanking bins provide lower significance detections and, in conjunction with the X-ray image and radio polarization, can be used to constrain rotating vector model solutions for the pulsar geometry.

\end{abstract}


\section{Introduction} \label{sec:intro}

PSR B1509$-$58 (=PSR J1513$-$5809) is a young ($\tau=1600$y), energetic (${\dot E}=1.7\times 10^{37}{\rm erg\,s^{-1}}$), high field ($B_s=1.5\times 10^{13}$\,G) pulsar embedded in the supernova remnant RCW89/G320.4$-$1.2/\msh15 \citep{1981MNRAS.195...89C}. The relativistic particles and fields produced by this pulsar power a bright X-ray pulsar wind nebula (PWN), whose spectacular {\it Chandra} X-ray Observatory ({\it CXO}) image has earned the moniker `The Cosmic Hand' or `The Hand of God'. This structure and the surrounding supernova remnant are detected from radio \citep{2016sros.confE..53L} to TeV \citep{2005A&A...435L..17A} energies with complex morphology, often complementary at different energy bands. At a distance $d\approx 5$\,kpc the $32^\prime$ diameter radio shell spans 47\,pc. The PWN's non-thermal X-ray emission extends $\sim 8^\prime$ from the pulsar, making the PWN complex $\sim\,4\times$ larger in angle and $\sim 10\times$ larger in size than the famous Crab PWN. \msh15 shares a `torus+ jet' morphology with the Crab, with a $\sim 10^{\prime\prime}$ sub-luminous X-ray zone around the pulsar representing the pre-termination shock flow \citep{2009PASJ...61..129Y}. The two bright X-ray arcs wrapping the northern side of the pulsar may represent the distorted equatorial torus of the shocked PWN or may represent field lines wrapped around the termination shock by ram pressure or backflow in the surrounding PWN (Figure 1). To the south along the torus axis is a prominent ridge of X-ray emission extending at least $5^\prime$, often referred to as a `jet' \citep{2002ApJ...569..878G}. To the northwest, non-thermal X-ray ridges form the `thumb' and `fingers' of the hand. The fingers extend to a region of softer thermal X-ray emission to the north.

The $P_s=150$\,ms pulsations are detected in the radio \citep{1982ApJ...262L..31M}, X-ray \citep{1982ApJ...256L..45S} and $\gamma$-ray \citep{2010ApJ...714..927A} bands. The pulsed spectral energy distribution (SED) is actually quite soft for a $\gamma$-ray pulsar, peaking at $\sim 10$\, MeV, which may be associated with its relatively large dipole field. As for most young gamma-ray pulsars, the high energy emission lags the radio peak, here by $\Delta \phi \approx 0.3$. At X-ray energies, the peak has two overlapping peaks, with separation $\delta \phi \approx 0.2$ \citep{2003A&A...400.1013D}.   

Existing polarization information on this system is limited. While the supernova shell itself is quite bright in the radio, the non-thermal emission is radio-faint. Radio observations with the Australia Telescope Compact Array (ATCA) at 3\, cm and 6\, cm have detected significant linear polarization, especially in the torus-like arcs \citep[][Zhang, et al.\,in prep.]{2016sros.confE..53L}. Here the inferred magnetic field follows the arcs as they wrap around the pulsar. To the south, this polarized radio emission brackets the X-ray jet. The X-ray jet fills a cavity in the radio emission, with little or no radio flux apparent, as also noted by \citet{2002ApJ...569..878G}. To the north, radio emission seems to follow the thumb and finger structures but is rather faint for reliable polarization maps. Like many young energetic pulsars, B1509-58 shows high linear polarization in the radio \cite{2001AJ....122.2001C,2015MNRAS.446.3356R}. 
From the {\it ROSAT} X-ray PWN structure \cite{1997MNRAS.284..335B} qualitatively estimated the viewing angle $i > 70^\circ$, although a somewhat smaller value is indicated by {\it CXO} data (Figure \,\ref{fig:Center}). 
There is a claim of a pulsar phase averaged optical polarization of degree $PD\sim$10.4\% by \cite{2000ASPC..202..315W}, but the measurement is compromised by a bright field star, and lacking any error bar or position angle (PA) estimate, this measurement needs to be confirmed.

Here we report on the first measurements of X-ray polarization from this complex, with robust detections in both the pulsar and the surrounding PWN, and describe how these results constrain the system geometry.

\begin{figure}[t]
\includegraphics[width=0.50\textwidth]{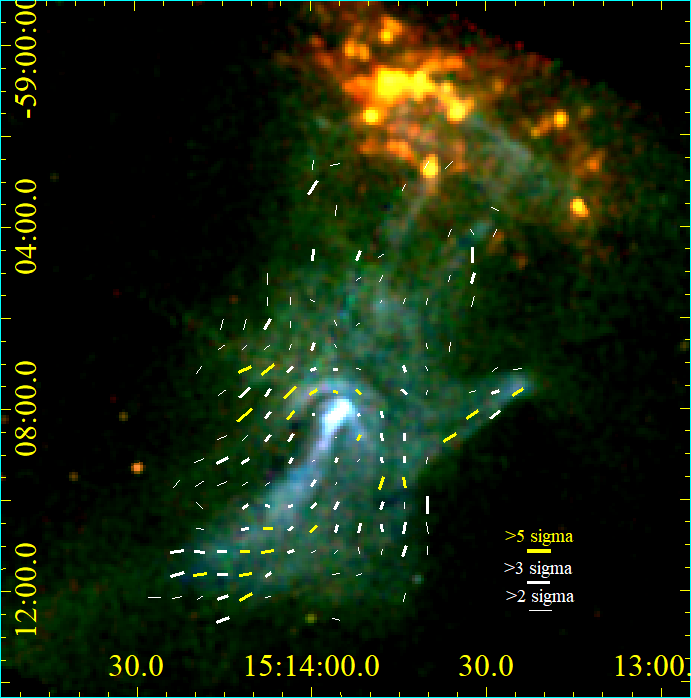}
 \caption{Overview of the \msh15 complex, the `Cosmic Hand'. The background energy-coded {\it CXO} image ($0.5-1.2$\,keV=red, $1.2-2$\,keV=green, $2-7$\,keV=blue) shows low energy (red/yellow) emission from thermal gas in G320.4$-$1.2, and harder non-thermal emission from the PWN. Superposed are {\it IXPE} 2-8\,keV {\tt PCUBE}-derived bars showing the polarization degree (PD, bar length is for PD=1) and projected magnetic field direction, on a 30$^{\prime\prime}$ grid. Yellow bars show $>5\sigma$ polarization detections, thick white bars $>3\sigma$ and thin white bars $>2\sigma$. In general the magnetic field appears to follow the thumb, fingers and other linear structures (see Fig.\,2 for region labels). Particularly strong polarization is associated with the arc, especially to the northeast of the pulsar and with the far southeast end of the jet. We have used a flux cut to trim anomalous vectors near the field-of-view edge. }
\label{fig:CXO+Pol}
\end{figure}
\section{{\it IXPE} Observations of PSR B1509-58/\msh15}

The Imaging X-ray Polarimetry Explorer ({\it IXPE}), the first mission devoted to spatially-resolved polarization measurements in X-rays \citep{weisskopf_imaging_2022}, was successfully launched on December 9 2001. {\it IXPE} observed \msh15 on 2-16 September 2022, 14-21 February 2023 and 13-19 March 2023 for a total of $\sim 1.5$\,Ms livetime. Data were extracted and analysed according to standard procedures: \texttt{HEASOFT} 6.30.1 \citep{2014ascl.soft08004N}
was used to perform barycenter corrections using the DE421 JPL ephemeris. \texttt{ixpeobssim} V30.2.2 \citep{baldini_ixpeobssim_2022} was used to do energy calibration, detector WCS correction, bad aspect-ratio corrections, and all further analysis, including phase folding at the pulsar ephemeris. 

Background events were cleaned from the data following the procedure of \citet{di_marco_handling_2023}. The residual instrumental background was modeled from 1.5Ms of cleaned {\it IXPE} source-free exposure in the fields of several high-latitude sources (MGC$-$5$-$23$-$16, 1ES 0299+200, PG 1553, PSR B0540$-$69 and IC 4329A). \msh15 lies close to the Galactic ridge so some contribution from background X-rays is expected as well. However, it covers most of the {\it IXPE} field of view so we cannot extract a local background spectrum directly from {\it IXPE}. Instead we use {\it CXO} observations to compute the background flux, passed through the {\it IXPE} instrument response, south of the thumb, finding a count rate $\sim 1.1\times$ the instrumental background, and so we increase the background spectrum surface brightness by this factor. This unpolarized background surface brightness ($8.9\times 10^{-8}$cnts/arcsec$^2$/s/det, 2-5.5\,keV; $1.07\times 10^{-7}$cnts/arcsec$^2$/s/det, 2-8\,keV) is scaled and subtracted from the flux of each aperture. 

Since \citet{2006ApJ...640..929D} have noted temporal variations in the fine structure of the PWN, especially in knots near the pulsar, but also in the jet feature, we collected a contemporaneous 28\,ks {\it CXO} observation of the PWN (ObsIDs 23540, 27448) to have a current high resolution image for comparison. 

Figure~1 gives an overview of the {\it IXPE} polarization measurements superimposed on an energy-coded image from archival {\it CXO} exposures (ObsIDs 0754, 3833, 5534, 5535, 6116, 6117 -- 204\,ks livetime total). Here we show the projected magnetic field direction (orthogonal to the Electric Vector Position Angle, EVPA) measured on a $30^{\prime\prime}$ grid, comparable to the resolution of the {\it IXPE} PSF. Complex polarization features extend throughout the nebula. 

\begin{figure}[t]
\includegraphics[width=0.45\textwidth]{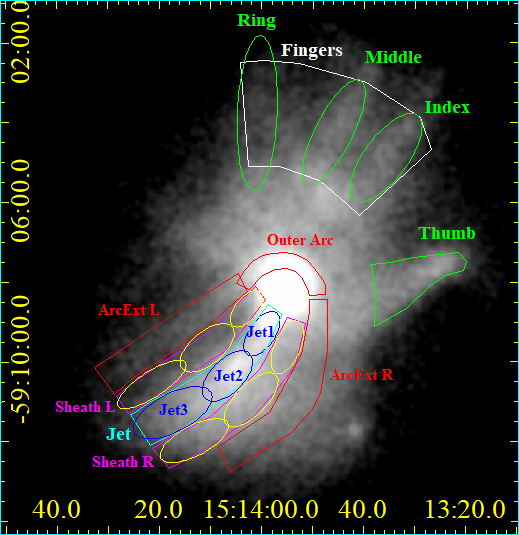}
 \caption{Morphological regions of interest on the {\it IXPE} combined DU1-3 2-8\,keV image. The Hand's `wrist' includes the bright jet and flanking filament structures, divided here into several regions.}
\label{fig:IXPE+reg}
\end{figure}

We can use the deep archival data to define nebula regions of interest (Figure 2 and Table 1). This allows us to discuss the polarization properties of extended regions too faint for high S/N mapping. The 2022 {\it CXO} image does show small departures from the archival morphology; most are changes in the shock structure near the pulsar, within the central {\it IXPE} resolution element, although there are also small changes in the jet $1-2^\prime$ from the pulsar. None affect the locations of our extended regions. In the region near the pulsar, {\it CXO}-maps show an inner arc, not resolved by {\it IXPE} (Figure 3). Strong, high significance polarization follows the outer arc, as also seen in the 6cm radio maps. The arc magnetic field structure extends, with field lines parallel to the jet, in left and right `arc extensions'. It is more faintly visible in the sheath regions flanking the X-ray bright jet. The field also clearly follows the curve of the thumb region. All of these features are also discernible in the radio. In addition we measure fields paralleling the `finger' structures -- in the radio, these are lost to bright emission from shock in the SNR shell. The shell interaction produces the low energy X-ray emission appearing in red and yellow to the north in Figure~1. This thermal emission, near the edge of the {\it IXPE} field-of-view, is unpolarized.  

The general pattern of polarization in Fig.\,1 is as expected, with the magnetic field lines following the filamentary nebula structure. The highest fractional polarizations, in the outer arc, thumb, and end of the jet, reach $PD\sim 70\%$ (after background subtraction). By integrating over the regions of Figure~2, we also see that the magnetic field is aligned with the thin `finger' structures, but with substantial background from the thermal emission at the fingertips, we suspect that the $PD$ in these regions is underestimated. 

\begin{table*}[t]
\centering
\caption{{\it CXO} spectral measurements and {\it IXPE} {\tt PCUBE} Polarizations for the regions of Figure \ref{fig:IXPE+reg}.}
\begin{tabular}{lrrrrrrrrr}
\toprule
Region & Flux &$\Gamma_X$ & $B_{Eq}$ & Q & U & Q, U err & PD & $\psi$ &Sig \\
& $0.5-8\,$keV& & $\mu$G & & & & & $\circ$ & \\
\midrule
Fingers &26.98  &$2.35\pm 0.01$ &  21  &$-0.119$ &$0.114$& 0.022& $0.165\pm0.022$& $68.0 \pm 3.8$ & 7.5\\
Index & 5.64 &$2.22\pm 0.01$ &  24     &$-0.216$ &$0.065$& 0.044& $0.225\pm0.044$& $81.7 \pm 5.6$ & 5.1\\
Middle & 6.65  &$2.34\pm 0.01$ &  29   &$-0.055$ &$0.183$& 0.043& $0.191\pm0.043$& $53.3 \pm 6.5$ & 4.4\\
Ring &  4.95  & $2.26\pm 0.02$ &  22   &$-0.177$ &$0.121$& 0.050& $0.214\pm0.050$& $72.8 \pm 6.7$ & 4.3\\
Thumb$^\dagger$&6.15&$1.92\pm 0.01$&16 &$0.313$  &$-0.085$& 0.038& $0.324\pm0.038$& $-7.6 \pm 3.4$  & 8.4\\
Outer Arc$^\dagger$&6.38&$1.87\pm 0.01$&27&$0.348$&$-0.020$&0.026&$0.348\pm0.026$& $-1.7 \pm 2.1$ &13.5\\
ArcExt L&  4.10 & $1.93\pm 0.01$ &  22 &$0.084$ &$0.412$& 0.043  &$0.421\pm0.043$& $39.2 \pm 2.9$ & 9.7\\
ArcExt R&  9.12 & $1.84\pm 0.01$ &  16 &$-0.109$ &$0.075$& 0.027 &$0.133\pm0.027$& $72.7 \pm 5.9$ &  4.9\\
Jet&  15.81  & $1.69\pm 0.01$ &  18    &$0.238$ &$0.089$& 0.021 &$0.254\pm0.021$& $10.2 \pm 2.3$ & 12.2\\
Sheath L&  $3.47$ & $1.77\pm 0.02$& 16 &$-0.001$&$0.142$& 0.034 &$0.142\pm0.034$& $45.2 \pm 6.9$ & 4.2\\
Sheath R& 9.68 & $1.80\pm 0.01$ &  20  &$0.092$ &$0.256$& 0.025 &$0.272\pm0.025$& $35.1 \pm 2.6$ &11.0\\
Jet-Sheath$^*$&    &  &                     &$0.602$ &$-0.146$&0.075 &$0.620\pm0.075$& $-6.8 \pm 3.4$ & 8.3\\

J1        & 4.58 &$1.64\pm0.01$&  25   &$0.012$ &$0.173$& 0.037&$0.173\pm0.037$& $43.1 \pm 6.1$ & 4.7\\
J2        & 6.54 &$1.74\pm0.01$&  19   &$0.276$ &$0.034$& 0.035&$0.278\pm0.035$& $3.5  \pm 3.6$ & 8.0\\
J3        & 4.59 &$1.67\pm0.01$&  16   &$0.466$ &$0.053$& 0.041&$0.469\pm0.041$& $3.2  \pm 2.5$ &11.5\\

J1-Sh1$^*$&    & &                          &$0.281$ &$0.123$& 0.109&$0.307\pm0.109$& $11.8 \pm 10.1$ & 2.8\\
J2-Sh2$^*$&   & &                           &$0.526$ &$-0.293$& 0.090&$0.602\pm0.090$& $-14.6 \pm 4.3$ & 6.7\\
J3-Sh3$^*$&   &  &                          &$0.819$ &$-0.145$& 0.159&$0.831\pm0.159$& $-5.0 \pm 5.3$ & 5.2\\
\bottomrule
\end{tabular}
\tablecomments{$\Gamma_X$ -- photon index, $0.3-8\,$keV unabsorbed fluxes in $10^{-12}{\rm erg\,cm^{-2}s^{-1}}$, (spectral fits include a common PhAbs model $N_H=8.96\pm 0.04 \times 10^{21} {\rm cm^{-2}}$ absorption), Equiparition fields are for $\sigma=\phi=1$. $PD$ -- polarization fraction, $\psi$ -- EVPA. \\$^*$average of flanking background subtracted for Jet and Jet sub-regions. $\dagger$ (Outer arc, Thumb) polarization measured along an arc; $\psi=0^\circ$ represents a $B$ field oriented along the arc.}
\label{tab:xspec}
\end{table*}

The most unusual feature is the X-ray bright hard spectrum `jet', which is essentially invisible in the radio, implying a low-energy cutoff in the jet electron spectrum. This may be an intrinsic cut-off in the injected electron spectrum or the result of limited time available for cooling in the rapid jet flow. We also note that the overall polarization level is low at the base of the jet region. Interestingly, the weak polarization that we do see appears to be at substantial angle to that of the bracketing nebula. These are, of course, 3-D structures so it seems likely that the jet zone is viewed through a plasma emitting like the `sheath' zones to either side. If we subtract the average Stokes I, Q, and U of this sheath (scaled for area) we do indeed see jet polarization increase to $PD > 60\%$, with the measured EVPA implying an average magnetic field at angles up to 50$^\circ$ from the jet axis.

\begin{figure}[h]
\includegraphics[width=0.4475\textwidth]{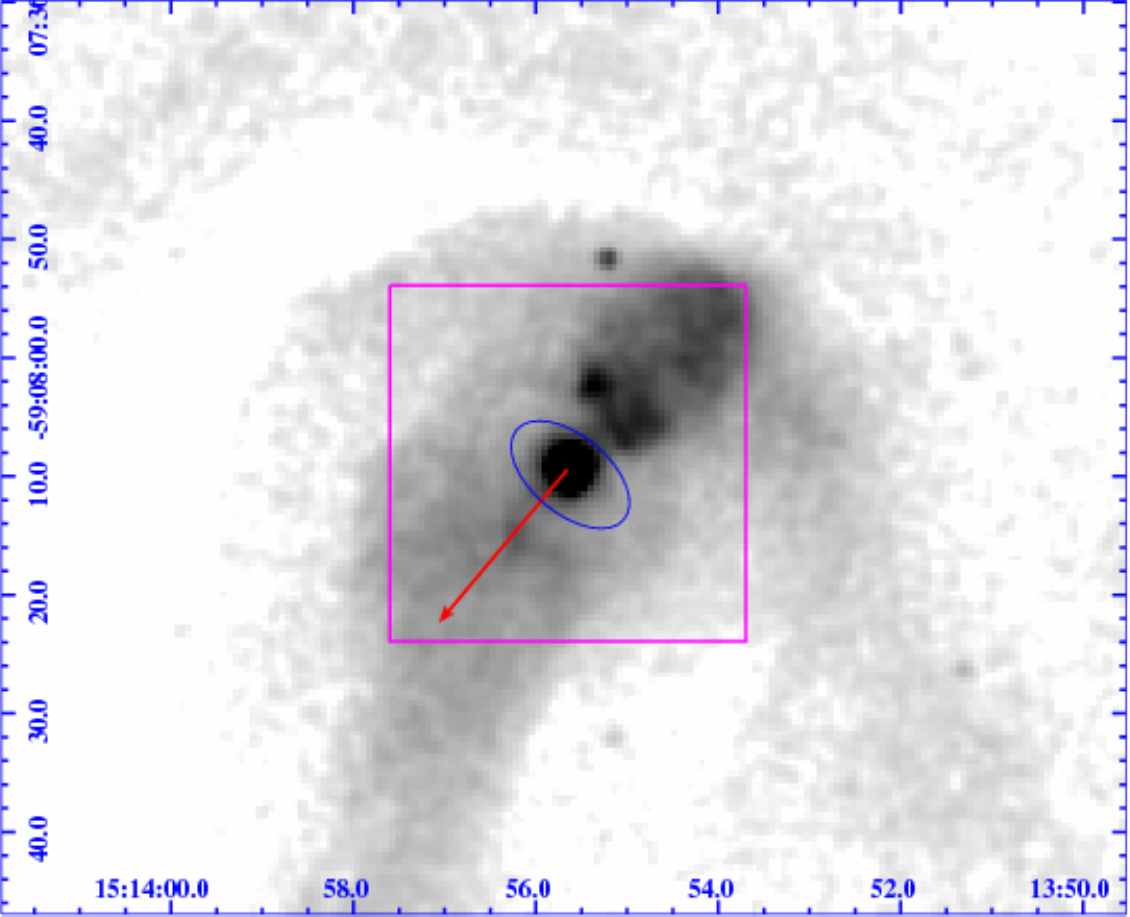}
\caption{Inner Region of \msh15 from $0.5-8$\,keV {\it CXO} data, centered on the pulsar and smoothed with a $1^{\prime\prime}$ Gaussian kernel. Note the sub-luminous zone (marked by an ellipse 2/3 of the zone size), the symmetry axis (red arrow), the bright polar outflow to the northwest and the fainter more diffuse outflow to the southeast. The magenta square marks the central pixel of the {\tt PCUBE} polarization map, Fig.~1 (approximating {\it IXPE}'s resolution); `simultaneous fit' pixels are half this size, Fig.~4. The `Inner Arc' passes through the northwest polar outflow, while the `Outer Arc' is partly visible to the north.}
\label{fig:Center}
\end{figure}

\section{Phase-Resolved Analysis of PSR B1509-58}

We obtained contemporaneous Parkes L-band radio observations (2023 2/7, 2/20, 2/26 and 3/1), and folded them with the same ephemeris used for the {\it IXPE} X-ray events to confirm that the $\Delta \phi=0.25$ radio-X-ray phase lag \citep{1991ApJ...383L..65K, 2001A&A...375..397C} remains valid. 

To extract the nebula map and the pulsar polarization we employ the `Simultaneous Fitting' technique of \citet{2023arXiv230608788W}. This uses the contemporaneous 2022 {\it CXO} image of the nebula, with the point source subtracted, to define the intensity (and local spectrum) of the extended emission at the {\it IXPE} observation epoch. For the pulsar point source contribution, we also rely on {\it CXO} data, using the ACIS-CC and HRC analysis of \citet{2017ApJ...838..156H} to define the light curve and phase-varying spectral index of the pulsar emission. The phase dependent pulsar and spatially dependent nebula spectra are folded through the {\it IXPE} response, using {\tt ixpeobsim}, to predict the {\it IXPE} counts as a function of position and phase. Note that PSR B1509$-$58 is relatively bright at minimum, at $\sim 4\%$ of its peak flux -- this means that the phase invariant DC emission contributes $\sim 11\%$ of the pulsed flux. To model faint PWN regions the uniform background must be included in the simultaneous fitting model; here we use the instrumental background, as local photon background is included in the {\it CXO}-derived flux.

Simultaneous fitting defines a set of spatial and phase bins, and uses the predicted {\it IXPE} counts from the nebula, background, and PSF-spread pulsar to define the expected PSR/PWN contributions to each bin. It then executes a global least-squares fit for the pulsar polarization at each phase and the phase-independent nebular polarization at each spatial pixel. Here we define a $13\times11$, $15^{\prime\prime}$ pixel grid centered on the  pulsar. We use $2-5.5$\,keV photons, to best isolate the pulsar polarization signal (which is slightly softer than the spectrum of the inner PWN) and simple `Moments Ellipticity' weights to quantify the accuracy of the polarization reconstruction of each event. Note that with different PSFs for each detector (as measured from ground calibration images) and different spacecraft orientations for each of the three {\it IXPE} pointings, we have nine measurements of the combined PSR/PWN polarization signal in each spatial and phase bin, all of which must be simultaneously fit. 

Because errors in the reconstructed photon conversion point are correlated with the reconstructed polarization vector, bright point sources (and any sharp flux gradient) will have a `halo' of polarization at scales less than the PSF FWHM, which can be corrected by an iterative estimate of this so-called polarization leakage \citep{2023A&A...672A..66B}. Here we apply the energy-dependent version of this correction, using the detailed ground-measured PSFs of the three telescope assemblies, as outlined in \citet{2023arXiv230608788W}. This correction is applied before the simultaneous fit extraction of the component polarizations. The correction makes modest ($<20\%$) amendments to the polarization degree in the inner few arcmin, especially associated with the relatively sharp arcs to the north of the pulsar. Also, when the spatial bins are smaller than the PSF FWHM and the counts/bin are low, anti-correlated fluctuations between adjacent pixels increase the scatter and error in fit $q$ and $u$. 
Here we mitigated this by smoothing the $q$ and $u$ maps by the PSFs for a decrease in fluctuations, at a cost of some spatial resolution. 

\begin{figure}[t]
\includegraphics[width=0.5\textwidth]{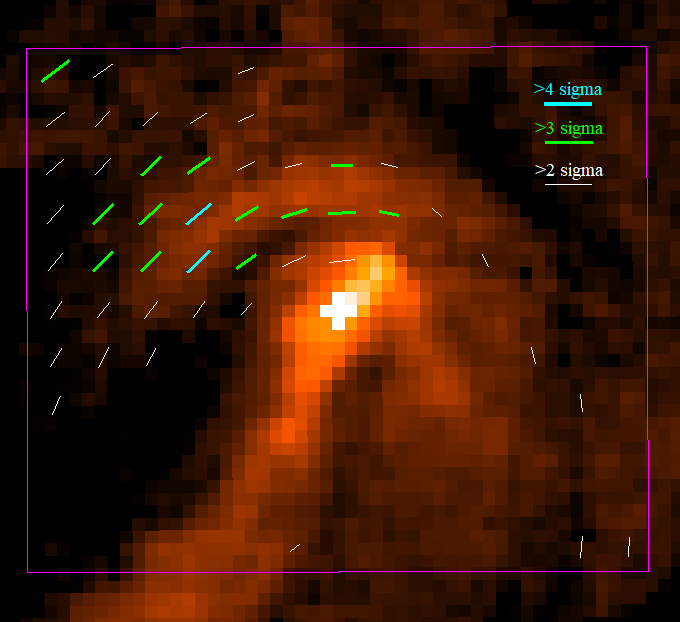}
 \caption{2-5.5\,keV polarization bars (orthogonal to the EVPA), as determined from simultaneous fitting inside the magenta region (cyan bars $>4\sigma$ significance, green bars $>3\sigma$, white bars $>2\sigma$; the labels at upper right show the bar length for PD=1). Here we employ a $15^{\prime\prime}$ grid to better resolve the pulsar from inner nebula structures. Background is the full-band {\it CXO} image. }
\label{fig:Simul}
\end{figure}

In Figure~5 we show the simultaneous fitting-derived pulsar X-ray EVPA estimates along with radio polarization measurements and the {\it IXPE} X-ray light curve for reference. Only one X-ray phase bin, near the center of the peak, is significant with a $PD$ of 17.5\% at $3.7\sigma$. The large pulse-minimum bin formally has a very high $PD \sim 1$ at low significance. However the {\tt PCUBE} analysis shows small total polarization in the central pixel -- simultaneous fitting evidently optimizes the central region fit by introducing some $q$ and $u$ into the faint minimum phase pulsar component and producing canceling PSR minimum and PWN polarizations in this phase bin. The other bins in the X-ray peak have low $PD=10-20\%$; at $\sim 2-3\sigma$ significance/bin there is no definitive polarization detection, although the EVPA values do assume an intriguing smooth sweep across the X-ray peak.

\section{Discussion}

The background-subtracted, leakage-corrected polarization map (Fig. \ref{fig:CXO+Pol}) has several $>5\sigma$ polarization regions. The most significant (Left Arc extension) pixel has a background-subtracted $PD=0.72\pm 0.08$. A few low-count pixels near the nebula edge have higher PD, the most extreme being in the Left Arc Extension, with $PD=0.87\pm 0.14$. Thus all pixels are consistent with $PD<0.75$ at the $1\sigma$ level.  Other highly polarized pixels are at the jet end ($PD=0.65\pm0.12$), the thumb base ($PD=0.66\pm0.11$) and the index finger ($PD=0.73\pm0.20$). The jet as a region is highly polarized toward its end with $PD=0.83 \pm 0.16$ at its far (J3) end, if one subtracts the adjacent sheath emission as a background. Thus, as also seen in the Vela PWN \citep{2022Natur.612..658X}, polarization approaches $PD=\Gamma_X/(\Gamma_X+2/3)$, the maximum allowed for synchrotron polarization at the observed X-ray photon index $\Gamma_X$ in a uniform magnetic field. For example, in the J3-Sh3 region, the maximum allowed value is $PD=0.72$; the observed polarization is $0.7\sigma$ above this value, consistent with a statistical fluctuation.

In the inner region, simultaneous fitting lets us map fields closer to the pulsar (Fig.\,\ref{fig:Simul}). Here again the strongest polarization follows the outer arc and the left arc extension, with a peak value of $PD=0.64\pm 0.14$. There is also a $PD=0.54\pm0.19$ polarized pixel of modest $2.8\sigma$ significance located on the ridge of the inner arc. This should be a pure nebula measurement, as it comes from the nebula portion of the simultaneous fit, generated with the contemporaneous {\it CXO}-defined structure, and has also been corrected for polarization leakage. However, at only $15^{\prime\prime}$ from the pulsar PSF peak, some concern about systematic effects persists. At both scales, polarization at the base of the jet is low.

Table 1 lists the average polarization degree and angle in the larger regions defined in Figure \ref{fig:IXPE+reg}. We can estimate the regions' magnetic field strength under the assumption of equipartition. For an optically thin region filled with relativistic electrons and magnetic field emitting synchrotron radiation the equipartition field is
\begin{equation}
\label{eq:syncB}
B_{\rm Eq} = 46 \left[ \frac{J_{\rm -20}(E_1,E_2) \sigma } {\phi}  \frac{C_{1.5-\Gamma}(E_m, E_M)}{C_{2-\Gamma}(E_1,E_2)}\right] ^{2/7} \mu G
\end{equation}
where 
\begin{equation}
C_q(x_1,x_2) = \frac{x_2^{q} - x_1^q}{q}.
\end{equation}
$J_{\rm -20}(E_1,E_2) = 4 \pi f_{X}(E_1, E_2) d^{2}/ V$ is the observed emissivity (in $10^{-20}$\,erg\,s$^{-1}$\,cm$^{-3}$, between $E_1$\,keV and $E_2$\,keV), $\sigma_B=w_B/w_e$ is the magnetization parameter, $\phi$ the filling factor, and $E_m$ and $E_M$ the minimum and maximum energies, in keV, of the synchrotron spectrum with photon index $\Gamma$. We assume that the structures are cylindrical, with diameter set to the observed region width. We list the derived equipartition fields in Table \ref{tab:xspec} for $\sigma=\phi=1$, $E_m=0.01\,\rm{keV}$ and $E_M=10\,\rm{keV}$.

\msh15 is complex but a few trends can be extracted from Table \ref{tab:xspec}. First, the fingers region is notably softer than the bulk of the PWN. This may, in part be due to contamination by the soft thermal emission to the north. However, the `Thumb' with $\Gamma=1.92$ is free of the thermal emission but still somewhat softer than the outer Arc. The hardest feature is, of course, the `jet' as seen in the color image (Fig.\,\ref{fig:CXO+Pol}). This suggests that this feature contains the freshest electron population and that the outer features have suffered some synchrotron burn-off. Indeed the jet may represent a site of $e^\pm$ re-acceleration and the low polarization at its base may be, in part, due to magnetic turbulence and dissipation there. The spectral trends are broadly consistent with those found by \citet{2014ApJ...793...90A}. These authors, using NuSTAR, infer a nebula-averaged spectral break at $\sim 6$\,keV. Thus, the average {\it CXO} spectral indices shown here should not resolve a full $\Delta \Gamma=0.5$ cooling break. 

Our equipartition field estimates are of course subject to the uncertain 3-D geometry and $\phi$ fill factor. There does seem to be a trend of higher fields to the north, which may be associated with compression from interaction with G320.4$-$1.2. We also note that the equipartition field strength appears to decrease along the jet, although the field becomes more uniform, as shown by the $PD$ increase as one moves away from the pulsar.

The field orientations for the morphologial regions support the pattern in Fig.\,\ref{fig:CXO+Pol}, with the arc and thumb fields well aligned with the curved ridges. The mean `Jet' field is oriented $\sim 25-35^\circ$ from the surrounding `Sheath' regions. If we imagine that these are 3D structures, with the Sheath surrounding the jet, we can subtract the mean sheath flux, to find that the offset angle increases to $\sim 40-50^\circ$ and the residual polarization is quite high at $PD=62\pm 8$\%. The jet field is not fully transverse to the jet axis, but the significant orthogonal component might implicate a helical structure. We subdivided the jet into three regions, finding a strongly increasing polarization as one moves downstream. The B orientations do not show a smooth trend, even after subtracting flanking sheath fluxes. The brightest mid jet region, however has the largest angle to the local jet axis at $\approx 50^\circ$. 

\begin{figure}[h!!]
\includegraphics[width=0.52\textwidth,trim={0.7cm 5cm 0cm 3cm}, clip]{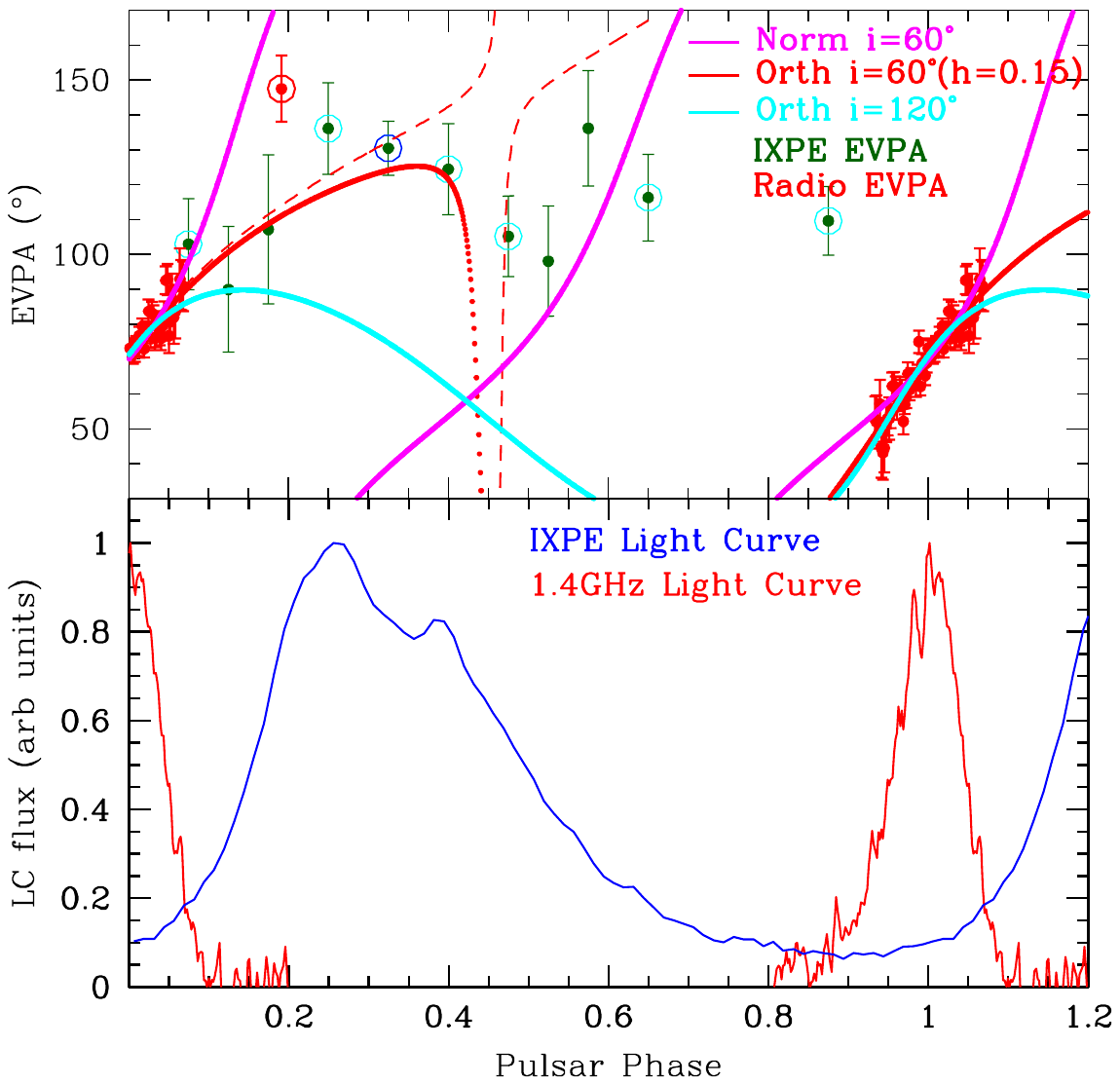}
 \caption{{\it IXPE} Top: phase bin PA estimates (green; $q$ and $u$ values available as data behind figure). The blue-circled bin at pulse center is individually significant ($3.7\sigma$); cyan-circled points are at $2-3\sigma$. The radio PAs (red) can be fit (excluding the circled point) with the normal mode  (magenta) and orthogonal mode (red) $i\sim 60^\circ$ models. For the $i\sim 120^\circ$ model only the orthogonal mode solution is tenable. Note that the $i\sim 60^\circ$ models better accommodate the late radio point and the X-ray data. Somewhat better match can be found for the orthogonal mode model when the X-ray emission is at higher altitude, with modest altitude $h$ (dashed curve).
 Bottom: {\it IXPE} pulsar light curve (blue), Parked 1.4GHz light curve (red).
 }
\label{fig:EVPA}
\end{figure}

Pulsar polarization can also be related to the PWN geometry. Examining the {\it CXO}-measured fine structure in the inner nebula, we see a general symmetry axis at $\psi = 140 \pm 5^\circ$. PWN features are best described by tangential views of structure in the MHD flow, as described by \citet{2003MNRAS.344L..93K}, but the geometrical inferences from a torus-jet picture  with cylindrical symmetry are robust. As for the Crab, there is a sub-luminous zone surrounding the pulsar, first described for \msh15 by \citet{2009PASJ...61..129Y}, which marks the equatorial flow prior to the termination shock (marked in Fig.~3 by an ellipse). For the Crab this zone is bracketed by the inner ring and wisps. The Crab wisps are brighter to the northwest, and if interpreted as due to Doppler boosting in mildly relativistic post-shock flow \citep{2008ApJ...673..411N} this determines the 3-D orientation of the spin axis. For \msh15 the zone has no bright edge and the Doppler boosting is not obvious. So while the ellipticity of the zone constrains the spin axis inclination to the Earth line-of-sight, both $i= 60\pm2^\circ$ with the southeast axis out of the plane of the sky (since the `jet' to the southeast would then approach us, we call this the `Jet' solution) and $i= 120\pm2^\circ$ with that axis into the plane of the sky (the `C-Jet' solution) are viable. One might interpret the blob of polar emission to the northwest in Fig.\,3 as a Doppler boosted `jet'. However, it is diffuse and is more likely outflow analogous to the dome of PWN emission northwest of the Crab, rather than a collimated relativistic jet flow. 

We can compare this geometry with that inferred from radio pulsar polarization measurements. Figure 5 shows Parkes 1.4GHz EVPA values, referenced to infinite frequency for a rotation measure of $RM=216.0\,{\rm rad/m^2}$ \citep{2006MNRAS.368.1856J}, where phase bins with linear polarization detected with $>2.5\sigma$ significance are plotted. Traditionally one fits the EVPA data $\psi(\phi)$  to the rotating vector model \citep[RVM,][]{1969ApL.....3..225R}, which can be generalized to include the effect of Doppler boosting of the rotating emission point at height $h=r/R_{\rm LC}$ as \citep[e.g.,][]{2020A&A...641A.166P}
\begin{multline*}
    {\rm tan} (\psi-\psi_0)=\\
    \qquad \frac 
    {{\rm sin}\theta\,{\rm sin}(\phi-\phi_0)+h[{\rm sin}i\,{\rm sin}\theta +{\rm cos}i\,{\rm cos}\theta\,{\rm cos}(\phi-\phi_0)]   }
    {{\rm cos}i\,{\rm sin}\theta\,{\rm cos}(\phi-\phi_0) - {\rm sin}i\,{\rm cos}\theta - h {\rm cos}\theta\,{\rm sin}(\phi-\phi_0)}
\end{multline*}
where $h \approx 0$ for the low altitude radio emission. Here $i$ is the inclination of the spin axis to the line of sight, $\theta$ is the angle between the magnetic and spin axes and the magnetic axis passes closest to the line of sight at $\phi=\phi_0$ with impact parameter $\beta=i-\theta$ and EVPA $\psi_0$. Note that the sign of the denominator addresses the `$\psi$ convention problem' \citep{2001ApJ...553..341E}. With the limited radio phase coverage, a simple $h=0$ fit to the radio data is not particularly constraining \citep{2015MNRAS.446.3367R}, but if we impose the prior constraints on $\psi_0$ \citep[which is either along or orthogonal to the projected spin axis at phase $\phi_0$,][]{2005MNRAS.364.1397J} and $i$ (two options, above) from the X-ray image, we obtain fits with well constrained parameters and small covariance. In Table \ref{tab:rfit} we show the Markov Chain Monte Carlo fit parameters for the three viable options (the normal mode orientation with $i \approx 120^\circ$ provides no acceptable fit to the radio data). Both orthogonal mode solutions provide very good fits. The one normal mode solution is worse, but with a p-value of $0.025$, still acceptable. If the X-ray image constraints are relaxed, the best-fit solutions remain stable, although errors of course increase and there is substantial $\phi_0 - \psi_0$ covariance.

\begin{table}[t]
\centering
\caption{MCMC Radio RVM fits -- see  Figure \ref{fig:EVPA}.}
\begin{tabular}{lrrr}
\toprule
Parameter &Norm Jet  &Orth Jet & Orth C-Jet \\
\midrule
$i (^\circ)$ & $58.0\pm1.9$ &$60.0\pm 1.9$ & $119.7\pm 2.0$ \\
$\theta(^\circ)$& $87.6\pm3.8$ &$123.0\pm 2.7$ & $148.5\pm 1.6$\\
$\phi_0 (^\circ)$ & $48.9\pm3.3$ &$-22.6\pm 3.0$ & $-18.1\pm 2.4$ \\
$\Psi_0 (^\circ)$ & $138.1\pm2.7$ &$49.9\pm 2.9$ & $52.8\pm 2.7$  \\
$\beta (^\circ) $       &  -28.7 & -63.0 & -28.8 \\
$\chi^2$/63 DoF &1.37  & 1.13 &1.19 \\
\bottomrule
\end{tabular}
\tablecomments{Priors from X-ray (Fig.\,\ref{fig:Center}) -- $\psi_0=140\pm3^\circ$ (Norm), $50\pm3^\circ$ (Orth), $i=60\pm2^\circ$ (Jet), $i=120\pm2^\circ$ (C-Jet).}
\label{tab:rfit}
\end{table}

The {\it IXPE} polarization data (and the late phase radio point) can help us distinguish between RVM models. The $i\approx 120^\circ$ RVM model cannot explain these points as the model EVPA is far off. The $i\approx 60^\circ$ models have EVPA increasing past the radio peak, and so more plausibly account for these data. In fact, for these models the post radio EVPA increases slightly for higher altitude emission; the orthogonal $i\approx 60^\circ$ model can match the {\it IXPE} EVPA and approach the late phase radio point if their emission is from higher altitude with $h>0$. A fit to the X-ray data formally gives $h \approx 0.15\pm 0.05$, but there are multiple minima and large departures at late phases. Note that the $i\approx 60^\circ$ orthogonal model sweep is slowest near the significant {\it IXPE} detection. For this case some loss of polarization signal might be attributed to sweep in the surrounding bins. Non-zero $h$ does not help the normal mode model, as it already has EVPA larger than that of the late phase points. 

The $i\approx 60^\circ$ orthogonal RVM model has the minimum $\chi^2$ but there is a major peculiarity: the radio peak appears when the associated magnetic pole sweeps $|\beta|=|i-\theta|=63^\circ$ from the Earth line-of-sight, while the opposite pole, sweeping $3^\circ$ away at $\phi= 0.44$ shows no radio emission. In contrast the other two models have large, but less extreme $\beta \approx -29^\circ$. Of these the Normal mode model has the radio pulse leading the radio axis by a substantial $49^\circ$, while the orthogonal solution would have the radio pulse trailing the magnetic axis. Thus, no solution is ideal and all require a very large, partly filled radio beam. Additional significant X-ray EVPAs would certainly help the model discrimination, as would more late-phase radio measurements.

\section{Conclusions}

In sum, the {\it CXO}-measured X-ray morphology of the inner PWN does constrain the 3-D spin axis and helps select between otherwise viable RVM fits to the PSR B1509$-$58 radio polarization. With {\it IXPE} we also extract a single phase bin of pulsar X-ray polarization. This is plausibly interpreted as an extension of the radio polarization sweep, but with only one significant bin, it its difficult to make detailed model tests. Further {\it IXPE} observations could promote 2-3 more bins to 3$\sigma$ significance, but would probably require $\sim2$\,Ms of additional exposure.

We conclude by noting that the rich polarization structure of the \msh15 PWN reflects the interplay of axisymetric pulsar outflow and complex, possibly unstable interaction with the surrounding SNR. Although, unlike the Crab and Vela PWNe, toroidal symmetry does not dominate the polarization pattern, the polarization degree of \msh15 is similar to that found earlier by {\it IXPE} for the Crab and Vela: polarization is very high in parts of the hard spectrum emission regions, approaching the maximum $PD$ allowed for synchrotron emission \citep{2022Natur.612..658X}. This suggests that these portions of the PWN contain uniform fields with little turbulence. On the other hand, the base of the jet which may be re-accelerating particles has a low polarization and complex field geometry. It seems that if diffusive shock acceleration (DSA) energizes the PWN particles, then much of the radiation comes from uniform field zones separate from the acceleration sites. Alternatively a lower-turbulence mechanism, possibly associated with magnetic reconnection, may be involved. Full mapping of the field geometry requires higher resolution and sensitivity than {\it IXPE} can provide. But even the present data provide a visually striking polarization map of the Cosmic Hand's fields and some important challenges to MHD PWN models.

\bigskip
\facilities{ATCA, {\it CXO}, {\it IXPE}}

\section*{acknowledgments}
The Imaging X-ray Polarimetry Explorer ({\it IXPE}) is a joint US and Italian mission.  The US contribution is supported by the National Aeronautics and Space Administration (NASA) and led and managed by its Marshall Space Flight Center (MSFC), with industry partner Ball Aerospace (contract NNM15AA18C).  The Italian contribution is supported by the Italian Space Agency (Agenzia Spaziale Italiana, ASI) through contract ASI-OHBI-2017-12-I.0, agreements ASI-INAF-2017-12-H0 and ASI-INFN-2017.13-H0, and its Space Science Data Center (SSDC) with agreements ASI-INAF-2022-14-HH.0 and ASI-INFN 2021-43-HH.0, and by the Istituto Nazionale di Astrofisica (INAF) and the Istituto Nazionale di Fisica Nucleare (INFN) in Italy.  This research used data products provided by the {\it IXPE} Team (MSFC, SSDC, INAF, and INFN) and distributed with additional software tools by the High-Energy Astrophysics Science Archive Research Center (HEASARC), at NASA Goddard Space Flight Center (GSFC).  Funding for this work was provided in part by contract NNM17AA26C from the MSFC to Stanford and 80MSFC17C0012 to MIT in support of the {\it IXPE} project.  Support for this work was provided in part by the NASA through the Smithsonian Astrophysical Observatory (SAO)contract SV3-73016 to MIT for support of the {\it Chandra} X-Ray Center (CXC), which is operated by SAO for and on behalf of NASA under contract NAS8-03060. C.-Y. Ng and Y.-J. Yang are supported by a GRF grant of the Hong Kong Government under HKU 17305419. N.B. was supported by the INAF MiniGrant ``PWNnumpol - Numerical Studies of Pulsar Wind Nebulae in The Light of IXPE''.

\bibliography{references}
\bibliographystyle{aasjournal}

\end{document}